\documentclass[aps,twocolumn,showpacs]{revtex4}

\usepackage{graphicx}
\usepackage{amsmath}
\usepackage{amssymb}
\usepackage{bm}
\def\e{\varepsilon}
\def\be{\begin{eqnarray}}
\def\ee{\end{eqnarray}}

\begin{document}

\title{Dephasing in a quantum dot coupled to a quantum point contact}

\author{Tomosuke Aono}
\affiliation{%
  Department of Physics,
  Ben-Gurion University of the Negev,
  Beer-Sheva 84105, Israel}
\date{\today}
\pacs{%
03.65.Yz, 
72.15.Qm, 
73.23.-b, 
74.50.+r,  
}

\begin{abstract}
We investigate a dephasing mechanism in a quantum dot capacitively coupled to a quantum point contact.
We use a model which was  proposed to explain the 0.7 structure in point contacts, based on the presence of a quasi-bound state in a point contact.
The dephasing rate is examined in terms of charge fluctuations of electrons in the bound state.
We address a recent experiment by
Avinun-Kalish {\it et al.} [Phys. Rev. Lett. {\bf 92}, 156801 (2004)],
where a double peak structure appears in the suppressed conductance through the quantum dot.
We show that the two conducting channels induced by the bound state  are responsible for the peak structure. 

\end{abstract}

\maketitle

{\it Introduction.} 
Coherent transmission of electrons through 
a quantum dot (QD) has been investigated using an Aharonov-Bohm (AB) interferometer to understand phase coherent transport transport~\cite{Yacoby95,Schuster97}.
For this purpose, controlled dephasing experiments are essential.
In Refs.~\cite{Buks98,Sprinzak}, experiments were performed
using mesoscopic structures with QDs.
Ref.~\cite{Buks98} measured
the suppression of coherent transmission through a QD embedded in an AB ring.
A quantum point contact (QPC) is capacitively
coupled to a QD in the Coulomb blockade regime.
Adding an electron to the QD changes the transmission probability $T$
through the QPC by $\Delta T$.
When the source-drain voltage $V_{\rm QPC}$ through the QPC is finite,
there are the current fluctuations, {\it i.e.} the shot noise.
The QPC then induces dephasing in the QD.
The visibility of the AB interference pattern is
$1-\alpha$~\cite{Aleiner97,Levinson97,Buks98,Hackenbroich}
with $\alpha= \gamma/\Gamma$,
where $\Gamma$ is the level width of the QD and
\begin{eqnarray}\label{eq:dephasing-rate}
  \gamma = \frac{e V_{\rm QPC}}{8\pi} \frac{(\Delta T)^2}{T(1-T)}.
\end{eqnarray}

Recently a controlled dephasing experiment was investigated for
the QD in the Kondo regime~\cite{Avinun04}.
Preceding this experiment, 
Silva and Levit addressed the problem using the slave boson mean field theory~\cite{Silva}.
The conductance $G$ of the QD is suppressed by
$\Delta G = - G(V_{\rm QPC}=0) \alpha$
with
\begin{eqnarray}
  \alpha = \frac{\gamma}{T_{K}},
\label{eq:dephasing-Silva-Levit}
\end{eqnarray}
where $T_K$ is the Kondo temperature.
Later, Kang~\cite{Kang} using the $1/N$ expansion calculated that
\begin{eqnarray}
  \alpha =  \frac{\gamma^2}{\gamma^2+T_K^2}.
\label{eq:dephasing-Kang}
\end{eqnarray}
The experiment demonstrated several interesting features.
One of them is the magnitude of $\Delta G$.
It is about 30 times larger than Eq.~(\ref{eq:dephasing-Silva-Levit}).
Kang~\cite{Kang} showed that
the dephasing rate can be large when the QPC is geometrically asymmetric.
Another intriguing result is that 
$\Delta G$ shows a double peak structure as a function of $T$.
This result has not yet been addressed and
it is natural to associate the problem with
another intriguing feature of the QPC,
the 0.7 structure~\cite{Thomas96a,Thomas98}.

In many experiments,
the conductance $G_{\rm QPC}$ through the QPC shows an additional plateau near
$G_{\rm QPC}= 0.7 \times 2 e^2/h$ at zero magnetic field~\cite{Thomas96a,Thomas98,Kristensen,Reilly01,Cronenwett,Reilly02}.
Further experiments have been performed~\cite{Graham,Rokhinson,Morimoto06,Crook,Luscher}
to understand the features that cannot be explained by the conventional point contact model.
In parallel,
many theoretical studies have been made
using different models,
including an antiferromagnetic Wigner crystal~\cite{Matveev},
spontaneous subband splitting~\cite{Reilly05},
spin-dependent electron correlations~\cite{Bruus,Tokura,Cornaglia,Kindermann},
and numerical calculations using the density functional theory~\cite{Berggren,Hirose,Jaksch,Rejec,Ihnatsenka}.
Refs.~\cite{Hirose,Rejec,Ihnatsenka} demonstrated
the formation of a quasi-bound state in the QPC,
which is responsible for localized spins near the QPC.
Grounded in this finding,
a generalized Kondo model has been invoked to describe
transport properties through the QPC~\cite{Meir_2002}.
The bound state and the Coulomb interaction in the QPC cause
an additional plateau of $G$, which exhibits the Kondo effect.
Kondo physics has been observed at low temperature and voltage bias~\cite{Cronenwett}. 
In addition, recent experiments~\cite{Roche04,DiCarlo06} measured
the shot noise through the QPC as a function of magnetic field.
The results indicate two conducting channels 
with different transmission amplitudes.
Ref.~\cite{Golub06} showed that the model~\cite{Meir_2002} 
is consistent with the experimental results in Ref.~\cite{Roche04,DiCarlo06}.

In this paper,
we investigate a dephasing mechanism in  a QD using the generalized Kondo model~\cite{Meir_2002} in a QPC.
The dephasing rate is examined in terms of charge fluctuations of the quasi-bound state in a QPC.
The presence of the state in the QPC  accounts for a dephasing mechanism which is qualitatively different 
from the mechanism without the bound state.
The two conducting channels due to the bound state are responsible for a double peak structure of the dephasing rate, which is observed in Ref.~\cite{Avinun04}.

\begin{figure}[htbp] 
	  \centering
	  \includegraphics[width=7cm]{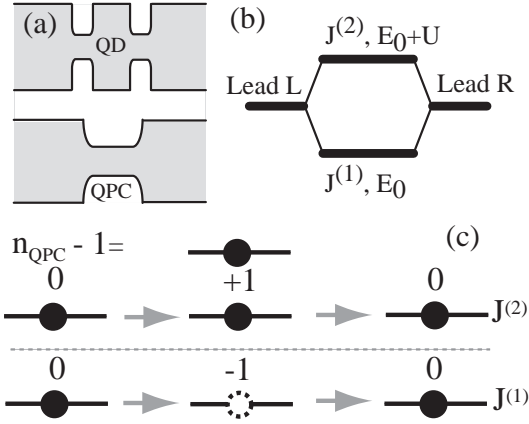}
	  \caption{%
	(a) Schematic view of a QD coupled to a QPC.
	(b) The two-channel model of the QPC.
	It consists of two conducting channels due to s-d coupling constants $J^{(i)} (i=1,2)$,
	which work as an Aharonov-Bohm interferometer in the QPC.
	(c) The charge fluctuations due to the transmission through the $J^{(1)}$ and
	$J^{(2)}$ channels. The numbers indicate $n_{\rm QPC}-1$.
	    \label{fig:qd}}
\end{figure}

{\it Model.} 
We consider a QD-QPC hybrid system as depicted in Fig.~\ref{fig:qd}(a).
The model Hamiltonian of the system consists of three parts,
$H_{\rm QPC}$, $H_{\rm QD}$, and $H_{\rm QPC-QD}$ as shown below.

The model Hamiltonian of the QPC proposed in Ref.~\cite{Meir_2002} is 
the generalized s-d model:
$H_{\rm QPC} = H_{\rm Lead} + H_{\rm sd}$ with
\begin{eqnarray}
  H_{\rm Lead} = \sum_{k\sigma \in L,R} \varepsilon_{k\sigma} \bar{c}_{k\sigma} c_{k\sigma},
\label{eq:H-qpc-lead}
\end{eqnarray}
\begin{eqnarray}
H_{\rm sd} &=& 
\sum_{k,k'\sigma \in L,R}
(J^{(1)}_{kk'}-J^{(2)}_{kk})
\bar{c}_{k\sigma}c_{k'\sigma}
\nonumber\\
 & +&2 \!\!\!\sum_ {k,k'\sigma\sigma' \in L,R}
(J^{(1)}_{kk'}+J^{(2)}_{kk'})
\bar{c}_{k\sigma}\vec{\sigma}_{\sigma\sigma'}
c_{k'\sigma'}\cdot \vec{S},
\label{eq:H-qpc-sd}
\end{eqnarray}
where
\begin{equation}
    J^{(i)}_{kk'}=
\frac{(-1)^{(i+1)}}{4}
\left[
\frac{V^{(i)}_{k}
V^{(i)}_{k'}}
 {\varepsilon_{k}-E^{(i)}}+
 \frac{V^{(i)}_{k}V^{(i)}_{k'}}
 {\varepsilon_{k'}-E^{(i)}}
\right],
\label{eq:def-J}
\end{equation}
and 
$\bar{c}_{k\sigma}$ creates an
electron with momentum $k$ and spin $\sigma$ in lead $L$ and $R$;
$E^{(1)}=E_0$ and
$E^{(2)}=E_0+U $ with
the energy level of local spin state $E_0$ and
the Coulomb energy $U$.
$\vec{S}$ is the local spin due to the localized state. 
We assume $J_{kk'}^{(i)} = J^{(i)}$.
The exponential increase of the couplings is modeled
by a Fermi function $f_{\rm FD}(x)=1/[1+\exp(x)]$, leading to 
the Fermi energy $E_F$ dependence
of $J^{(i)}$:
$
    J^{(i)}=\frac{(-1)^{(i+1)} (V^{(i)})^2 }{E_{F}-
    E^{(i)}}f_{\rm FD}(-E_{F}/\delta).
$

The Hamiltonian of the QD is 
the conventional Anderson model:
$ 
 H_{QD} = \sum_{k,\sigma, \atop \alpha=L,R}  \epsilon_k \bar{f}_{k \sigma \alpha} f_{k \sigma \alpha} +  \sum_{\sigma} \e_{0} \bar{d}_{\sigma} d_{\sigma} + 
  U_d n_{\uparrow} n_{\downarrow} + V_T \sum_{k,\sigma, \atop \alpha=L,R} \left[
 \bar{f}_{k \sigma \alpha} d_{\sigma} + {\rm h.c.} \right]$,
where $\bar{d}_{\sigma}$ creates an electron in the QD with spin
$\sigma$, while $\bar{f}_{k \sigma \alpha}$ creates an electron  with momentum $k$ and spin $\sigma$ in the lead $\alpha$ attached to the QD with the tunneling matrix element $V_T$;
$n_{\sigma} =\bar{d}_{\sigma} d_{\sigma}$, and $\e_0$ and $U_d$ are the energy level and the Coulomb energy in the QD, respectively.

The third part of the Hamiltonian, $H_{\rm QPC-QD}$ describes the interaction between the QPC and the QD.
Localized electrons in the QPC interact with the electrons in the QD:
\begin{eqnarray}
	H_{\rm QPC-QD} =  W n_{\rm QPC} \sum_{\sigma} n_{\sigma},
	\label{eq:time-dependent energy level}
\end{eqnarray}
where $n_{\rm QPC}$ is the number of localized electrons in the QPC,
and $W$ is the coupling constant.
The energy level in the QD is shifted by $H_{\rm QPC-QD}$:
$\e_0  \rightarrow \e_{0} + W n_{\rm QPC}$.

{\it Two channel induced dephasing.} 
The conductance through the QPC was calculated using second order perturbation theory~\cite{Meir_2002,Appelbaum}:
$G_{\rm QPC}=2 e^2 T/h$ with
\begin{eqnarray}\label{eq:T-perturbation}
    T &=& 4 \pi^2 (\rho_L \rho_R)^2
  \left(
  (J^{(1)}-J^{(2)})^2 + 3 (J^{(1)}+J_{r}^{(2)})^2 
\right)
\end{eqnarray}
and the density of states $\rho_{L} (\rho_R)$ in the left (right) leads.
We have introduced a renormalized coupling constant $J_{2;r} = 1/[4 \log(T/T_{K})]$
with the Kondo temperature $T_K = U \exp[-1/(4 \rho J_2)]$
that characterizes the Kondo effect in the QPC~\cite{Meir_2002}.
The right-hand side of Eq.~(\ref{eq:T-perturbation}) consists of three terms proportional to
$(J^{(1)})^{2}$, $(J^{(2)})^{2}$, and
$J^{(1)} J^{(2)}$.
This combination of the terms indicates
an AB interferometer picture with the $J^{(1)}$ and $J^{(2)}$ channels in the QPC as depicted in Fig.~\ref{fig:qd} (b).
Note that the appearance of the $J^{(1)}J^{(2)}$ term  is not peculiar to the s-d model (\ref{eq:H-qpc-sd}).
When multi channels involve electron transport, interference between them occurs.

The electron transport through the QPC induces fluctuations of $n_{\rm QPC}$
since it takes place via 
the co-tunneling processes described by 
$J^{(i)}$ in Eq.~(\ref{eq:def-J}).
If no current flows through the QPC, $n_{\rm QPC}=1$.
When electrons pass through the $J^{(1)}$ channel,
virtual excitations from $n_{\rm QPC}=1$ to $n_{\rm QPC}=0$ are involved,
while
when electrons pass through the $J^{(2)}$ channel,
excitations to $n_{\rm QPC}=0$ are involved.
These situations are depicted in Fig.~\ref{fig:qd} (c) with $n_{\rm QPC}-1$.
This change in $n_{\rm QPC}$ shifts $\e_0$ in the QD.
In this way,
the transmission of electrons through the QPC is monitored by electrons in the QD.
The current fluctuations (shot noise) through the QPC lead to fluctuations in $n_{\rm QPC}$ and eventually in $\e_0$.
It has been shown that the fluctuations of $\e_0$ due to the external environment lead to
dephasing in the QD, where the time evolution of $d_{\sigma}$ shows 
a exponential decay due to the fluctuations~\cite{Levinson97}.

Transport through the ``AB ring" in the QPC is monitored by the QD
through these charge fluctuations.
The terms proportional to $(J^{(1)})^2$ and $(J^{(2)})^2$ give
the transmission probability with $n_{\rm QPC}-1= \mp 1$, respectively.
These processes are monitored by the QD.
The $J^{(1)} J^{(2)}$ term, on the other hand,
describes the interference between the excited states with $n_{\rm QPC}-1 = \pm 1$.
This indicates that the term, compared to the $(J^{(1)})^2$ and $(J^{(2)})^2$ terms, involves smaller charge change in $n_{\rm QPC}$ after
an electron passes through the QPC.
In other words, the current fluctuations of the $(J^{(i)})^2$ terms contribute
to the dephasing in the QD while those of the $J^{(1)} J^{(2)}$ term can be negligible.

The dephasing rate $\gamma$ is then the sum of the dephasing rates of the two independent channels,
$(J^{(1)})^2$ and $(J^{(1)})^2$ terms.
In each channel,
we use the result of the previous theories~\cite{Buks98,Aleiner97,Levinson97,Hackenbroich} for a single channel QPC model.
The measured $\Delta T$ characterizes the interaction between the QPC and QD.
The total dephasing rate $\gamma$ is,
instead of Eq. (\ref{eq:dephasing-rate}),
\begin{eqnarray}
	\gamma = \frac{e V_{\rm QPC}}{8\pi} \left[\gamma_0(T_1) + \gamma_0(T_2) \right]
	\label{eq:dephasing-QPC-Kondo}
\end{eqnarray}
with
$
 \gamma_{0}(T) =  [\Delta T(T)]^2/T(1-T),
$
where $T_{i}$ is the transmission probability through the channel $J^{(i)}$:
\begin{eqnarray}
	T_{1/2} &=& 4 \pi^2 (\rho_L \rho_R)^2 (J_{1/2}^{2} + 3 J_{1/2r}^{2}).
	\label{eq:Ti-perturbation}
\end{eqnarray}
The common factor of $\Delta T(T)$ appears for both transmission channels.
This is because $\Delta T$ is measured by adding an electron to the QD,
and this affects both channels equally.

We calculate $\gamma$ in Eq.~(\ref{eq:dephasing-QPC-Kondo}) as a function of $T$ in Eq.~(\ref{eq:T-perturbation}).
We use a perturbative approach with
\begin{equation}
	\bar{T} = \frac{T}{1+T},
	\label{eq:T-effective}
	\end{equation}
	and
\begin{eqnarray}
	\bar{T}_{i} = \frac{T_i}{1 + T_{i}},
	\label{eq:Ti-effective}
\end{eqnarray}
in place of $T$ and $T_i$.
This corresponds to taking into account 
the perturbative corrections to $H_{\rm Lead}$ by $H_{\rm sd}$.
The current though the QPC is calculated in the following way.
We expand the Keldysh action
$
{\rm T_c} \exp (-i S_{\rm sd}),	
$
where $S_{\rm sd}$ is the action of the s-d interaction~(\ref{eq:H-qpc-sd}),
and the time order is taken along the Keldysh contour.
We expand the action up to the second order in $J^{(i)}$,
and reduce it to the bilinear form with
respect to conduction electron fields using Wick's theorem~\cite{decoupling}.
Then we have a non-interacting model without $H_{\rm sd}$ and with the renormalized action $\tilde{S}_0$ for the kinetic term of conduction electrons. Then the current through the QPC is calculated with $\tilde{S}_0$  and the current operator $
I = i e / \hbar [
 \sum_{k \in L, k' \in R, \sigma}
(J^{(1)} -J^{(2)}) \bar{c}_{k\sigma}c_{k'\sigma} 
+
2  \sum_ {k,\sigma \in L, \atop k' \sigma' \in R}
(J^{(1)}+J^{(2)})
\bar{c}_{k\sigma} \vec{\sigma}_{\sigma\sigma'} c_{k'\sigma'} \cdot \vec{S} ] + {\rm h.c.}.
$
The transmission probability is then given by Eq.~(\ref{eq:T-effective}).
If the renormalization of $S_0$ is disregarded, where $\tilde{S}_0 = S_0$ with
the action $S_0$ for Eq.~(\ref{eq:H-qpc-lead}), the transmission probability is given by
Eq.~(\ref{eq:T-perturbation}).
The origin of $T$ in the denominator of Eq.~(\ref{eq:T-effective}) is the s-d scattering in each lead,
while $T$ in the numerator is the scattering between two leads.
Since the s-d coupling constants are equal for both scatting processes,
the same factor of $T$ appears.
In a similar way, 
the transmission probability $T_{i}$ through the channel $J^{(i)}$
acquires the denominator, $1 + T_i$.

\begin{figure}[htbp]
  \centering
  \includegraphics[width=7cm]{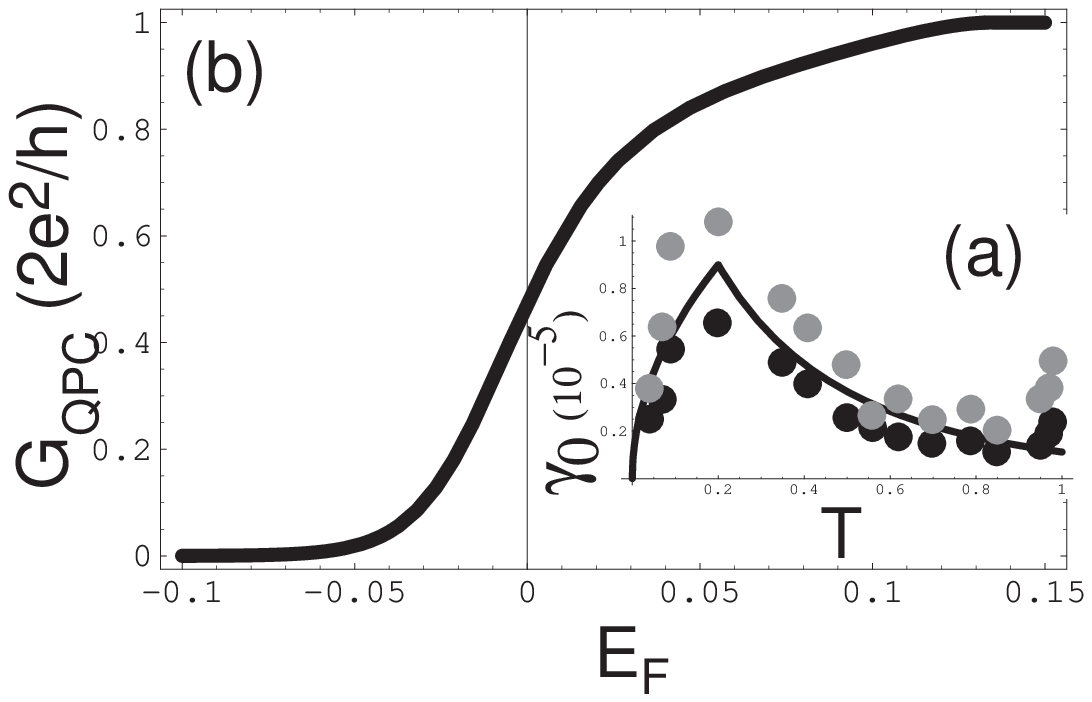}
  \includegraphics[width=7cm]{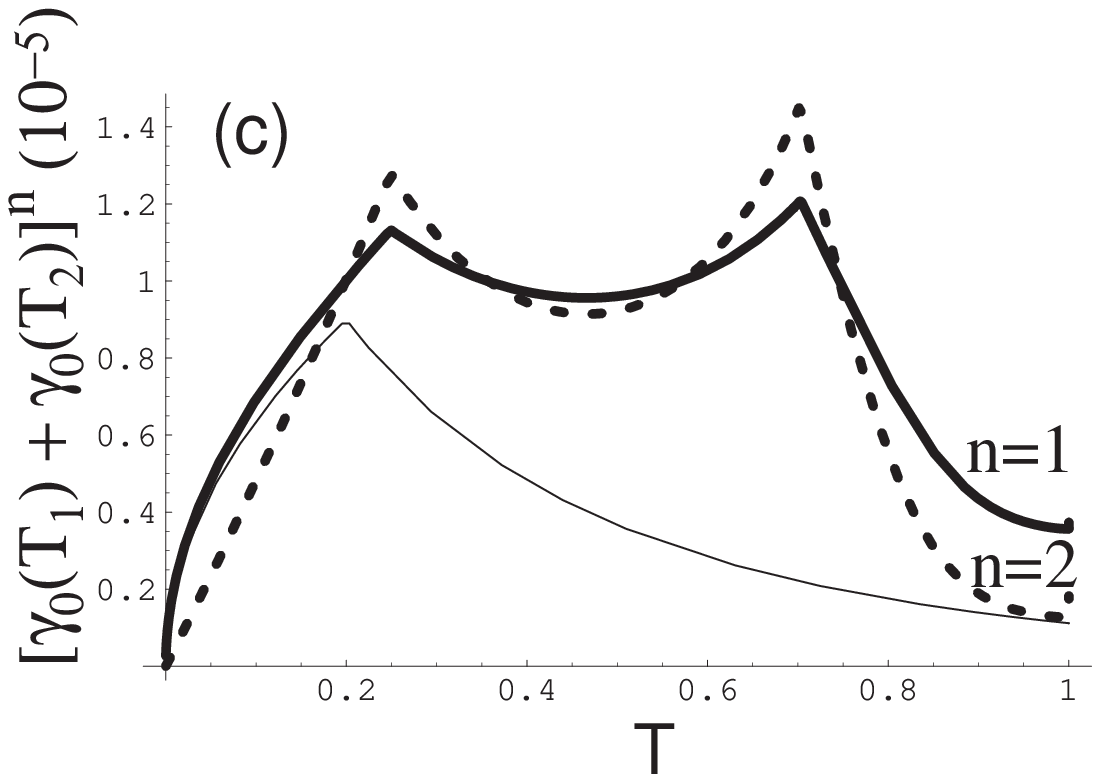}
  \caption{
    (a) $\gamma_0$ as a function of $T$. Symbols indicate the experimental data in Ref.~\cite{Avinun04}.
The solid line is the fitting curve.
    (b) $G_{\rm QPC}$ as a function of the Fermi energy $E_F$ of conduction electrons with
    $ \rho (V^{(1)})^2= 0.25, \rho (V^{(2)})^2 = 0.025, U = 1.5, \delta=0.02, k_B T = 0.001$.
	The unit of energy is $|E_0|$.
    (c) $[\gamma_0(T_1) + \gamma_0(T_2)]^{n}$ as a function of $T$. 
The thick solid line ($n=1$) is for Eq.~\ref{eq:dephasing-QPC-Kondo} while the thin line is for the conventional QPC model, and the broken line ($n=2$) for Kang's result~\cite{Kang}.
    \label{fig:results}}
\end{figure}

{\it Comparison with experiment.} 
We need to find the $T$ dependence of $\gamma_{0}$ from the experimental data. 
In Fig.~\ref{fig:results}(a),
symbols indicate the two sets of the experimental data in Ref.~\cite{Avinun04}.
To fit these data,
we use
$\gamma_{0}(T) = 0.9 \sqrt{T/0.2} \times 10^{-5}$ when $T < 0.2$ and
$\gamma_{0}(T) =  0.9 \exp(1-(T/0.2)^{0.7}) \times 10^{-5}$ when $T > 0.2$.
The plot is shown by the solid line in Fig.~\ref{fig:results}(a).
This choice of $\gamma_0$ reflects the fact that $\Delta T$ is a highly asymmetric function
with respect to $T$;
The maximum of $\Delta T$ is located at $T = 0.2$.
Other choices of $\gamma_0$ will give qualitatively similar results.

In the experiments,
the differential conductance through the QPC exhibited a zero bias anomaly (ZBA)  while
no clear sign of the 0.7 structure was observed.
In Ref.~\cite{Cronenwett}, a ZBA was observed,
which confirms that it originates from the Kondo effect.
The absence of a clear 0.7 structure does not contradict
the Kondo effect but rather it indicates that the effect is strong.
In Fig.~\ref{fig:results}(b),
$G_{\rm QPC}$ is plotted as a function of the Fermi energy $E_{\rm F}$ of conduction electrons
with $\rho (V^{(1)})^2/|E_0| = 0.25,\rho (V^{(2)})^2/|E_0| = 0.025$, and $U/|E_0| = 1.5$.
The parameters are chosen so that the QPC does not show a clear 0.7 structure in $G_{\rm QPC}$.

In Fig.~\ref{fig:results}(c),
$\gamma/(e V_{\rm QPC}/8\pi)= \gamma_{0}(T_1) + \gamma_{0}(T_2)$ is plotted as a function of $T$ by the thick solid line, while for comparison
$\gamma_{0}(T)$ for the conventional single channel QPC model is shown by the thin solid line.
A double peak structure of $\gamma$  appears as in the experiment,  in contrast to a single peak structure.
The peak positions are located at $T \sim 0.25$ and $T \sim 0.7$.
According to Ref.~\cite{Kang},
$\Delta G$ is given by Eq.~(\ref{eq:dephasing-Kang}).
It is proportional to $\gamma^2 \propto [ \gamma_0(T_1)+\gamma_0(T_2) ]^2$ when 
$\gamma \ll T_K$.
The broken line in Fig. \ref{fig:results} (c) shows the result for this case.
The double peak structure becomes more pronounced.

We should mention a consequence of the asymmetric line shape of $\Delta T$, which
questions the dephasing theory based on the conventional model of the QPC.
The dephasing rate is too small when $T \sim 0.7$ besides the absence of the extra peak.
If $\Delta T$ were symmetric, $\gamma_0(T)$ would be symmetric around $T=1/2$.
The difference between the experiment and theory was then quantitative, but not qualitative.
The experiment revealed an essential feature of the QPC.
For the two channel model used here, on the other hand,
this asymmetry helps to show the double peak structure.

{\it Discussion.} 
If the 0.7 structure of $G_{\rm QPC}$ is observed, the second peak near $T=0.7$ of $\gamma$ is sharper than
the one without the 0.7 structure.
This is because  the conductance is changed noticeably near $T=0.7$, 
and then the shot noise through the $J^{(2)}$ channel changes abruptly as well.

We did not address the amplitude of $\Delta G$.
As pointed out by Kang~\cite{Kang},
the asymmetrical structure of the QPC induces an larger dephasing rate in the experiment.
In this case, the dephasing rate depends on not only $\Delta T$ but also
on the change of the phase shift through the QPC, which requires additional information from
experiments, such as measurements in the device setup in Ref.~\cite{Sprinzak}.

In conclusion,
we have discussed the dephasing mechanism due to charge fluctuations of a
quasi-bound state in a quantum point contact.
The bound state is responsible for there being two transmission channels.
The dephasing rate is proportional to the sum of the transmission probability through 
these two channels.
This mechanism explains the double peak structure of the suppression rate of the conductance,
observed in a recent experiment~\cite{Avinun04}.
The result is qualitatively different from the rate without the bound state in the QPC.


The author acknowledges fruitful discussions with Y.~Meir, as well as valuable comments on the manuscript.
He also thanks to M.~Avinun-Kalish, Y.~Dubi, A.~Golub, T.~Rejec, and R.~S.~Tasgal for discussions.



\begin{thebibliography}{99}
\bibitem{decoupling}
It is useful to apply a unitary transformation for the conduction electrons,
introducing two linear combinations of $(c_{k \sigma}+ c_{k' \sigma'})/\sqrt{2}$ and 
$(c_{k \sigma}- c_{k' \sigma'})/\sqrt{2}$, where $k, \sigma \in L$ and $k',\sigma' \in R$.
Then the latter is decoupled from the s-d interaction.
See also, Ref.~\cite{Golub06}.
\end{thebibliography}
\end{document}